\documentclass[useAMS,usenatbib,usegraphicx,a4paper]{mn2e}
\newcommand {\aplt} {{\raise-.5ex\hbox{$\buildrel<\over\sim$}}} 
\newcommand{\objname}{2MASS\,J02132062+3648506}
\newcommand{\objnameshort}{2MASS\,J02132+3648}
 \title[A wide T dwarf companion to an active M dwarf binary]{2MASS~0213+3648~C: A wide T3 benchmark companion to an an active, old M dwarf binary}
 \author[N.R.\ Deacon et al.]{N.R.\ Deacon\thanks{E-mail:n.deacon2@herts.ac.uk}$^{1,2}$, E.A.\ Magnier$^3$, Michael C.\ Liu\thanks{Visiting Astronomer at the Infrared Telescope Facility, which is operated by the University of Hawaii under Cooperative Agreement no. NNX-08AE38A with the National Aeronautics and Space Administration, Science Mission Directorate, Planetary Astronomy Program.} $^{3}$, Joshua E. Schlieder $^{4,2}$,\newauthor
 Kimberly M. Aller$^3$, William M.J. Best$^3$, Brendan P. Bowler$^5$\thanks{McDonald Prize Fellow}, W.S. Burgett$^6$.\newauthor K.C. Chambers$^3$, P.W. Draper$^7$, H. Flewelling $^3$, K.W. Hodapp$^8$, N. Kaiser $^3$, \newauthor N. Metcalfe$^7$,W.E. Sweeney $^3$, R.J. Wainscoat $^3$, C. Waters$^3$\\
 $^1$Centre for Astrophysics Research, University of Hertfordshire, College Lane, Hatfield, AL10 9AB, UK\\
 $^2$Max Planck Institute for Astronomy, Konigstuhl 17, Heidelberg, 69117, Germany\\
 $^3$Institute for Astronomy, University of Hawaii at Manoa, 2680 Woodlawn Drive, Honolulu, HI, 96822, USA\\
 $^4$NASA Exoplanet Science Institute, Caltech, Pasadena, California, USA\\
 $^5$Department of Astronomy, The University of Texas at Austin, 2515 Speedway, Austin, TX 78712, USA\\
 $^6$Giant Magellan Telescope Observatory, USA\\
 $^7$Department of Physics, University of Durham, South Road, Durham DH1 3LE, UK\\
 $^8$Institute for Astronomy, University of Hawai'i, 640 North Aohoku Place, Hilo, HI 96720, USA}
 \begin{document}
 \date{}
 \pagerange{\pageref{firstpage}--\pageref{lastpage}} \pubyear{2015}
 \maketitle
 \label{firstpage}
 \begin{abstract}
 We present the discovery of a 360\,AU separation T3 companion to the tight (3.1\,AU) M4.5+M6.5 binary \objname. This companion was identified using Pan-STARRS\,1 data and, despite its relative proximity to the Sun (22.2$_{-4.0}^{+6.4}$\,pc; Pan-STARRS\,1 parallax) and brightness ($J$=15.3), appears to have been missed by previous studies due to its position near a diffraction spike in 2MASS. The close M~dwarf binary has active X-ray and H$\alpha$ emission and shows evidence for UV flares. The binary's weak {\it GALEX} UV emission and strong Na\,I 8200\,\AA\,Na absorption leads us to an age range of $\sim$1-10\,Gyr. Applying this age range to evolutionary models implies the wide companion has a mass of 0.063$\pm$0.009\,$M_{\odot}$. \objnameshort~C provides a relatively old benchmark close to the L/T transition and acts as a key, older comparison to the much younger early-T companions HN~Peg~B and GU~Psc~b.
 \end{abstract}
\section{Introduction}
Wide multiple systems are common in the solar neighbourhood, with $>$25\% of solar-type stars having a companion wider that 100\,AU \citep{Raghavan2010}. However the formation mechanism of these systems, especially those wider than $\sim$5,000\,AU, presents a challenge for models of star and brown dwarf formation. Current suggestions are that  such companions formed in situ, were captured in their birth cluster \citep{Kouwenhoven2010}, or result from the evolution of higher-order multiple systems \citep{Delgado-Donate2004,Umbreit2005,Reipurth2012}. This last theory predicts that wide binaries originate in tertiary systems and that dynamical interactions can drive these systems into being a close binary pair with a wider companion. Thus wide binary components should have a large, higher order multiplicity fraction (as seen in the literature review of substellar objects by \citealt{Burgasser2005} and observed for M dwarfs by \citealt{Law2010}) and that wide companions to close binaries should be more common than wide companions to single stars (seen by \citealt{Allen2012}).

Wide substellar companions in particular are also useful as test beds for evolutionary and atmospheric models. Such objects fall in to two categories; ``mass benchmarks'' \citep{Liu2008} where two of the system components are close enough to have an orbit that can be measured and thereby yield a dynamical mass, and ``age benchmarks", where the age of the primary is determined (typically from activity or rotation) and then applied to the secondary. Rare systems that are both age and mass benchmarks have both a higher mass primary for age determination and low-mass object with dynamical mass measurements (\citealt{Dupuy2009b,Crepp2012,Dupuy2014}).

The past decade has seen an explosion in the discovery of wide ultracool\footnote{Objects with spectral types M7 or later} companions to stars and brown dwarfs. Reviews of this population are presented in \cite{Faherty2010} and \cite{Deacon2014}.  Currently there are 22 known T dwarf companions with separations greater than 100\,AU, the 21 listed in the literature review of \cite{Deacon2014} plus the recently discovered companion to the exoplanet host star HIP~70849 \citep{Lodieu2014}. Of these, four (Gl 570 D, \citealt{Burgasser2000}; Wolf 1130 B, \citealt{Mace2013}; Ross 458 C, \citealt{Goldman2010}, and $\xi$ UMa E, \citealt{Wright2013}) lie in systems with higher-order stellar multiplicity. Relatively few known T dwarfs companions lie in young systems. HN~Peg~B is a T2.5 \citep{Luhman2007} companion to a $\sim$300\,Myr G0 dwarf. It is well fitted by the spectral standards with only a possibly weaker 1.25\,$\mu$m potassium doublet hinting at reduced gravity. By contrast the younger ($\sim$100\,Myr), lower mass T3.5 GU~Psc~b \citep{Naud2014} diverges significantly from the spectral standards in the $Y$, $J$ and $K$ bands. \cite{Naud2014} suggest that this may be due to a differing amount of collision induced absorption from molecular hydrogen or possibly due to an unresolved, cooler companion. These are the only T2.5--3.5 companions in the literature.

Here we present a wide separation T3 companion to a known, active M~dwarf binary system in the solar neighbourhood discovered during our search for ultracool companions using Pan-STARRS\,1. 

\section{Observations}
\subsection{Identification in Pan-STARRS\,1 data}
The Pan-STARRS\,1 telescope \citep{Kaiser2002} is a wide-field 1.8m telescope on Haleakala on Maui in the Hawaiian Islands and has recently completed its three and a half year science mission. Pan-STARRS\,1 data are reduced and calibrated photometrically and astrometrically by a process described by \cite{Magnier2006,Magnier2007,Magnier2013} and \cite{Schlafly2012}. The telescope undertook a series sky surveys including the 3$\pi$ Survey, covering the 30,000 sq.deg. north of $\delta=-30^{\circ}$ with two pairs of observations per year in each of five filters. These filters ($g_{P1}$, $r_{P1}$, $i_{P1}$, $z_{P1}$ and $y_{P1}$; \citealt{Tonry2012}) extend further into the red optical than SDSS, making Pan-STARRS\,1 appealing for studying late-type objects in the solar neighbourhood \citep{Deacon2011,Liu2011,Liu2013,Best2013,Best2015}. This combination of properties have already resulted in the identification of the largest number of ultracool companions to date, with a near-doubling of the late-M and L dwarf wide separation ($>$100\,AU) companion populations \citep{Deacon2014} as well as the discovery of two T dwarf companions \citep{Deacon2012a,Deacon2012} . 

\begin{figure*}
 \setlength{\unitlength}{1mm}
 \begin{picture}(100,100)
 \includegraphics{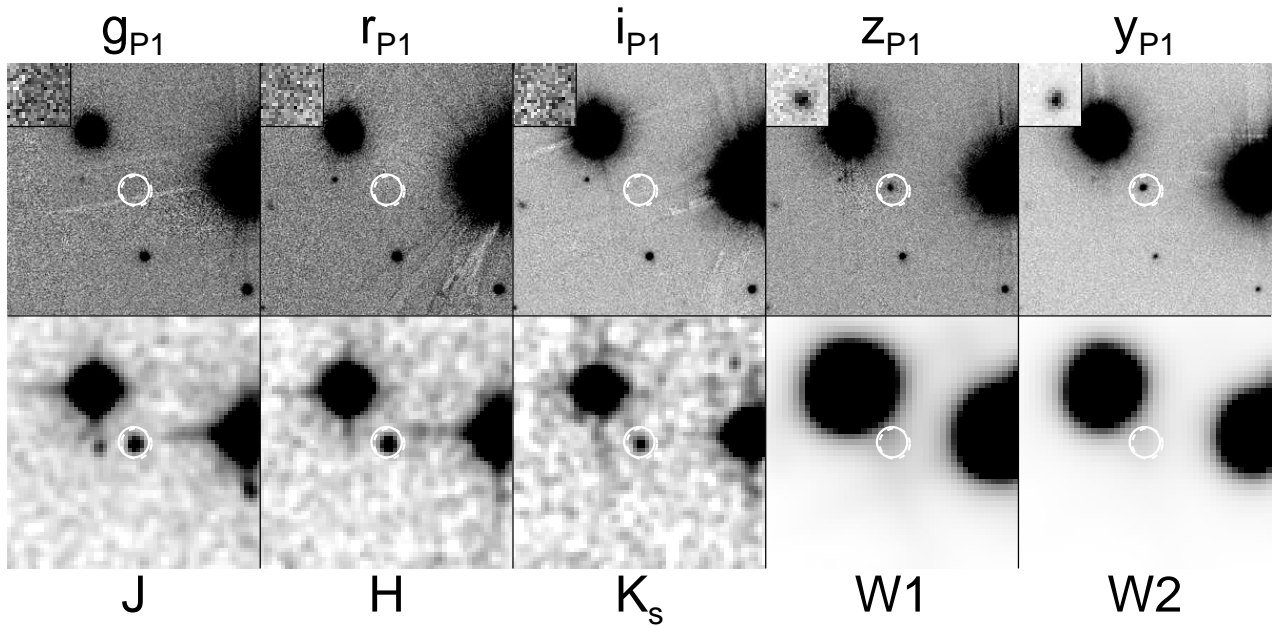}
 \end{picture}
 \caption[]{Discovery images of the wide T companion (centre) and its bright primary (to the North and East of the companion). Each image is 60$^{\prime\prime}$ across with North up and East left. The small cutouts in the PS1 images are 10$^{\prime\prime}$ across centred on the companion. Note the object is not detected in $W1$ and $W2$ possibly due to contamination from nearby bright stars. The solid circle marks the 2010.0 position and the dashed circle the 2MASS position. The PS1 image is a stack of images taken over the duration of the survey (2010--2013) so the object will not fall exactly at the 2010.0 position and may appear smeared due to proper motion.} 
  \label{sec_images}
 \end{figure*}

We searched the 3$\pi$ dataset processed by the first version of the Pan-STARRS\,1 pipeline to be run on the full survey area (aka. PV1) for wide companions to nearby stars. As in \cite{Deacon2014}, we included objects from various proper motion catalogues, but unlike this previous work we did not exclude nearby objects with low proper motion ($<$0.1\arcsec/yr). We followed the same selection process as \cite{Deacon2014} selecting only objects with proper motions measured over a $>$400\,day baseline, requiring a 5$\sigma$ overall significance and two individual detections in $y_{P1}$. This process included checking the 2MASS Reject Table as well as the main 2MASS database for near-infrared data. One candidate we identified was an apparent companion to a mid-M dwarf at $\sim$15\,pc \objname\,( aka PSO\,J33.3327+36.8105; hereafter \objnameshort). This object came into our input primary list from the nearby M~dwarf catalogue of \cite{Lepine2011}. \objnameshort\, was discovered as an M4.5 by \cite{Riaz2006} and identified as a close (0.217$"$) binary by \cite{Janson2012}, we refer to the unresolved M dwarf binary as \objnameshort~AB. From the contrast ratio ($\Delta i$ =2.16$\pm$0.15\,mag.) \cite{Janson2012} estimated a spectral type of M6.5 for the secondary component with the primary remaining an M4.5. The primary+secondary pair is saturated and unresolved in Pan-STARRS\,1 imaging. Our candidate common proper motion companion showed contaminated 2MASS photometry (flagged as c for confusion) from a nearby diffraction spike and did not appear in survey data taken by the Widefield Infrared Survey Explorer (WISE; \citealt{Wright2010}), neither in the main All-WISE table nor the All-WISE Reject Table \citep{Cutri2013}. This may be due to the proximity of two bright stars (including the primary) within half an arcminute. The flagged 2MASS detection and WISE non-detection are the likely reasons for this otherwise bright, nearby T dwarf remaining undiscovered until now. Figure~\ref{sec_images} shows the companion in a number of different Pan-STARRS\,1, 2MASS and WISE filters.

 \subsubsection{Probability of chance alignment}
 In order to estimate if our proposed companion and its primary is due to a chance alignment between unrelated objects, we used a modified version of the test presented in \cite{Lepine2007}. In the original test the positions in the input primary catalogue are offset by several degrees before the pairing between the primaries and with potential secondaries is carried out. This means the resulting data should only contain coincidental pairings. The pairings are then plotted on a proper motion difference vs. separation plot with a maximum separation of 10 arcminutes. Here we take a slightly different approach as we also have a trigonometric parallax for the companion (Section \ref{ps1ast_sec}). Hence we assign a quantity which we call "astrometric offset". This is a quadrature sum of the proper motion differences in each axis and the parallax difference each divided by the error on that quantity. Hence
 
 \begin{equation}
 n_{\sigma}^2 = \frac{(\mu_{\alpha 1} - \mu_{\alpha 2})^2}{(\sigma_{\mu \alpha 1}^2 + \sigma_{\mu \alpha 2}^2)} + \frac{(\mu_{\delta 1} - \mu_{\delta 2})^2}{(\sigma_{\mu \delta 1}^2 + \sigma_{\mu \delta 2}^2)} + \frac{(\pi_{1} - \pi_{2})^2}{(\sigma_{\pi 1}^2 + \sigma_{\pi 2}^2)}
 \end{equation}
 
\begin{figure}
 \setlength{\unitlength}{1mm}
 \begin{picture}(100,120)
 \includegraphics{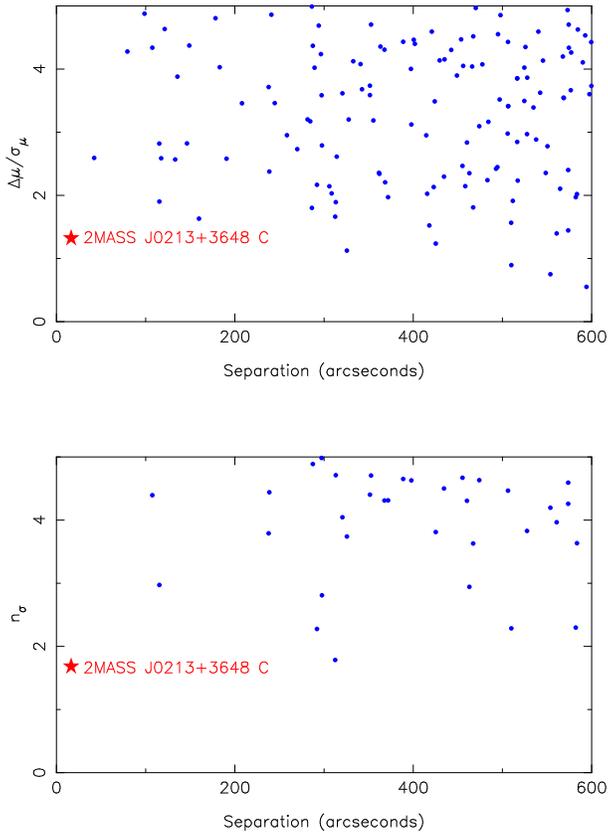}
 \end{picture}
 \caption[]{Comparing our companion with a sample coincident pairings generated by searching for late-type ($z_{P1}-y_{P1}>0.8$ companions around all M dwarfs in the catalogue of \cite{Lepine2011} with photometric distances within 20\,pc. The positions of these M dwarfs were offset by two degrees so only coincident pairings (blue dots) should be included. {\bf Top:} The y-axis shows the quadrature sum of the difference in proper motion between each pair in terms of the number of standard deviations {\bf Bottom:} The y-axis shows the quadrature sum of the differences in proper motion and parallax between each pair in terms of the number of standard deviations. The inclusion of parallax data clearly reduces the contamination from background stars.} 
  \label{shifted}
 \end{figure}

 We drew primaries with measured photometric distances within 20\,pc from the list of 8889 (7464 within the Pan-STARRS\,1 survey area) bright M dwarfs presented in \cite{Lepine2011}. These photometric parallaxes were used in the calculation of the astrometric offset. No lower proper motion limit was set on top of the 40\,mas/yr lower limit in our input primary list.  We then offset the positions of these object by two degrees in Right Ascension and carried out a pairing process with late-type ($z_{P1}-y_{P1}>0.8$ ) objects in the Pan-STARRS\,1 database. We made a conservative cut on the $\chi_{\nu}^2$ (the $\chi^2$ per degree of freedom) statistic of the astrometric points around the Pan-STARRS\,1 database parallax fit limiting it the be $<$10 and limited ourselves to objects detected in eight or more Pan-STARRS\,1 images (this latter cut is identical to one used in \citealt{Deacon2014}). The astrometry used for our pair were the proper motion and photometric parallax for the primary from \cite{Lepine2011} and our own proper motion and parallax for the secondary (calculated in section~\ref{ps1ast_sec}). Our results are shown in Figure~\ref{shifted} with plots showing $n_{\sigma}$ both with and with our the parallax term. Clearly the pairing of \objnameshort~AB  with our proposed companion is distinct from the coincident pairing locus. As there are no stars in our coincident pairing test with similar distances and astrometric offsets to \objnameshort we are unable to calculate a formal chance alignment probability. The inclusion of parallax data also significantly reduces the number of chance alignment pairings.
 
 \subsection{UK Infrared Telescope photometry}
 As the photometry from 2MASS was strongly affected by a diffraction spike, we obtained UKIRT/WFCAM \citep{Casali2007} observations of \objnameshort~C on UT 2015 July the 31$^{st}$. These observations were reduced using the WFCAM reduction pipeline \citep{Irwin2004,Hodgkin2009} by the Cambridge Astronomical Survey Unit. The resulting photometry is in Table~\ref{prop}.
 
\subsection{NASA Infrared Telescope Facility spectroscopy}
Our candidate companion was observed on UT October the 5$^{th}$ 2015 with the recently upgraded SpeX spectrograph \citep{Rayner2003} on the NASA Infrared Telescope Facility. Conditions were non-photometric with periods of high humidity and seeing of 0.6--1.0$''$. The observation was made at an airmass of 1.06 with three pairs of 50\,s observations nodded in an ABBA pattern. The low-resolution prism mode was used along with the 0.8$''$ slit giving a spectral resolution of R=75, with the slit aligned to the parallactic angle to minimise slit losses due to atmospheric dispersion. An A0V standard star (HD 22859) was observed contemporaneously at a similar airmass to the target. The spectrum was reduced using version 4.0 of the SpeXtool software package \citep{Cushing2004, Vacca2003}. The final spectrum for \objnameshort~C is shown in Figure~\ref{sec_spectrum} and the analysis of this spectrum is discussed in Section~\ref{spec_an}.

\begin{figure*}
 \setlength{\unitlength}{1mm}
 \begin{picture}(100,100)
 \includegraphics{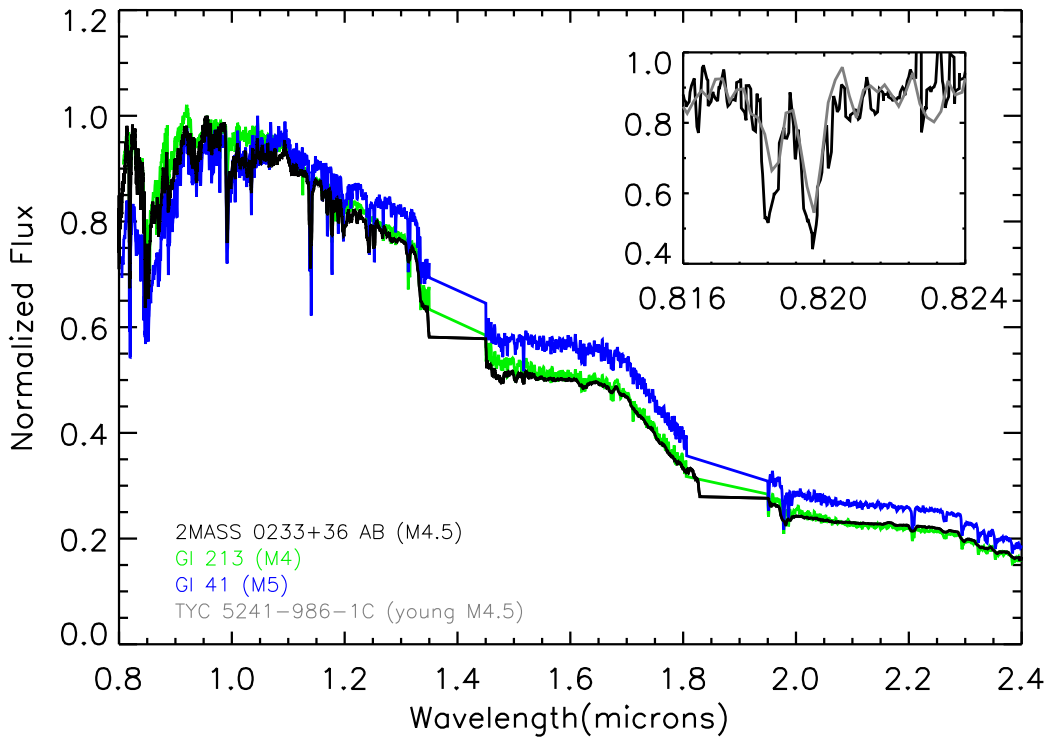}
 \end{picture}
 \caption[]{The SpeX/IRTF spectrum of \protect\objnameshort~AB. Two comparison spectra of M4 and M5 standards from \protect\cite{Kirkpatrick2010} are shown in the main plot. The spectra are normalised at 0.9\,$\mu$m. The inset shows the 8200\,\AA~feature compared to the 20\,Myr to $\sim$125\,Myr old M4.5 TYC 5241-986-1~C \protect\cite{Deacon2013}.} 
  \label{prispec}
 \end{figure*}

The unresolved primary \objnameshort~A/B was observed on UT September 25$^{th}$ 2015 with SpeX. Conditions were mostly clear with seeing of 0.5-0.8\arcsec. The 0.3\arcsec slit was used along with the cross-dispersed SXD mode yielding a spectral  resolution of R=2000. The observation was made at an airmass of 1.05 and consisted of four pairs of 14.8\,s observations nodded in an ABBA pattern. The reduction technique described above for the companion was also used but with the comparison star HD 19600. The spectrum is shown in Figure~\ref{prispec}. We discuss the use of the 8200\,\AA~Sodium doublet in this spectrum in section \ref{prim_age}

\subsection{Calar Alto 2.2\,m spectroscopy}
\objnameshort~AB was observed under poor conditions on the 11$^{th}$ of November 2014 UT using the Calar Alto Fiber-Fed Echelle spectrograph (CAFE, \citealt{Aceituno2013}) on the 2.2\,m telescope at Calar Alto Observatory in Almeria, Spain. A single 750\,s exposure was obtained under clear skies with high, variable humidity and poor seeing ($>$2\arcsec). Along with the science target, two radial velocity standards from \cite{Nidever2002} were observed for internal calibration checks, GJ~908 and HD~26794. Both standards were observed under slightly better conditions with single exposures of 400s and 800s, respectively. We also obtained a set of calibration observations prior to \objnameshort~AB and the standards that included bias frames, continuum lamp flats, and ThAr lamp frames for wavelength calibration. The data were reduced using the observatory pipeline described in \cite{Aceituno2013}, which bias corrects, flat fields and wavelength calibrates the data and applies a rough flux calibration. 

The reduced spectrum of \objnameshort~AB has a signal-to-noise-ratio (SNR) of $\sim$3 per pixel, the RV standards have SNRs of $\sim$25. To measure our target's radial velocity, we selected a region of the spectrum spanning 6600--6640\,\AA\, that was relatively free of telluric lines, had some strong stellar absorption lines, and was in the part of the spectrum with the highest SNR. This 40\AA\, region was flattened by dividing by the continuum polynomial fit and reduced from a resolution of $\sim$65000 to $\sim$48000 by convolving with a Gaussian kernel. This last step was taken to match the resolution of the late-type RV templates used for cross-correlation (CC) observed using FEROS on the ESO/MPG 2.2m telescope at La Silla. A barycentric velocity correction was also applied to the CAFE spectra. Our initial radial velocities for our RV standards were discrepant from the values in \cite{Nidever2002} by $-$58\,km/s. We were unable to find the source of this shift so can only correct for it post-hoc. This results in values of $-71.8\pm1.0$\,km/s compared to the literature value of $-71.3\pm0.1$\,km/s for GJ~908 and $57.0\pm1.0$\,km/s compared to $56.6\pm0.1$\,km/s for HD~26794. This process and correction yielded a radial velocity of 1.5$\pm$1.4\,km/s for \objnameshort~AB although the low signal to noise meant that the cross correlation power has a poor value of 20\%. The error in the radial velocity is the quadrature sum of the measured error on a Gaussian fit to the cross correlation peak and a systematic error of 1\,km/s introduced by the use of observed RV templates. There is also the additional factor of the RV amplitude induced by the secondary, this is approximately 2\,km/s.

\subsection{Pan-STARRS\,1 astrometry}
\label{ps1ast_sec}
Pan-STARRS\,1 has observed each field in the 3$\pi$ survey repeatedly over several years, allowing us
to measure the parallax and
proper motions of \objnameshort~C. We were unable to calculate a parallax or proper motion for \objnameshort~AB due to saturation. The Pan-STARRS\,1 astrometric analysis of \objnameshort~C consists of an initial series of transformations to remove distortions caused by the Pan-STARRS\,1 optical system and camera. An iterative astrometric correction process is then used to determine the conversion from chip to celestial coordinates, minimising the scatter between objectsÕ positions on each exposure. This process results in an astrometric systematic floor of $\approx$ 10 - 20 milliarcseconds.

The in-database measurement of parallax (0.080\arcsec$\pm$0.010) and proper motion does not yet
include a robust outlier rejection scheme. All 11 PS1 measurements spanning 3 years, were included in this analysis. We have found that a careful assessment of
possible outliers, and the impact of the outlier rejection, is
necessary to have confidence in the parallax and proper motion fits.
While we are working to incorporate these checks in the automatic
analysis, we have used an external calculation for \objnameshort~C
for the purpose of this article.

\begin{figure}
 \setlength{\unitlength}{1mm}
 \begin{picture}(100,80)
 \includegraphics{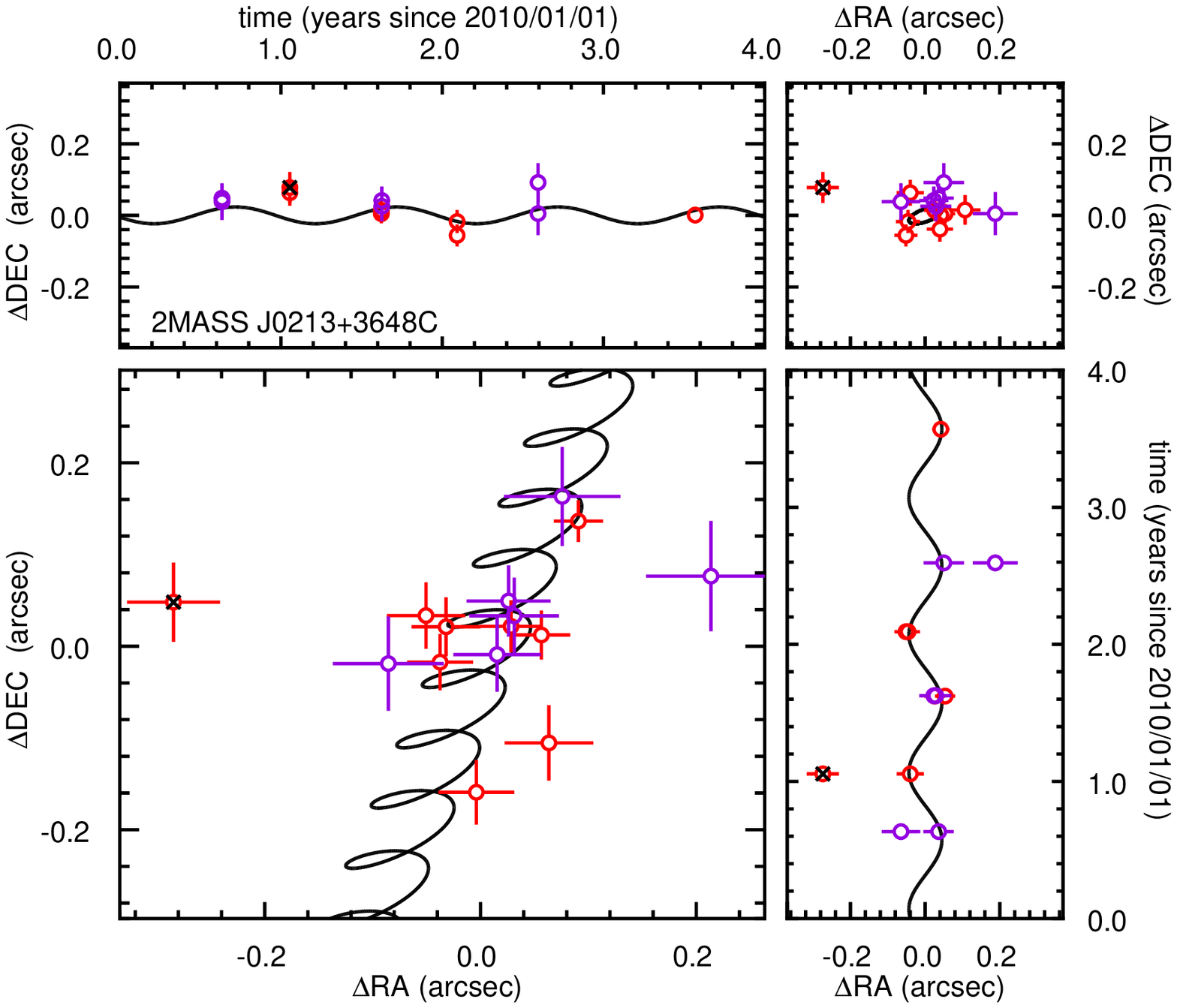}
 \end{picture}
 \caption[]{{\bf Lower left:} The
relative positions of the Pan-STARRS\,1 $z_{P1}$ (blue) and $y_{P1}$ (red)
measurements for \objnameshort~C. The best fitted path on the sky is drawn in black.  The
single point with a black X is suspect and excluded from the final fit.
{\bf Upper left:} Residual Declination positions as a function of
time (MJD) after the proper-motion fit has been subtracted; symbols
as above.  {\bf Lower right:} Same as Upper-left for Right
Ascension.  {\bf Upper right:} Residual Declination vs Right
Ascension after the proper-motion fit has been subtracted; symbols
as above.} 
  \label{par_data}
 \end{figure}

\begin{figure}
 \setlength{\unitlength}{1mm}
 \begin{picture}(100,80)
 \includegraphics{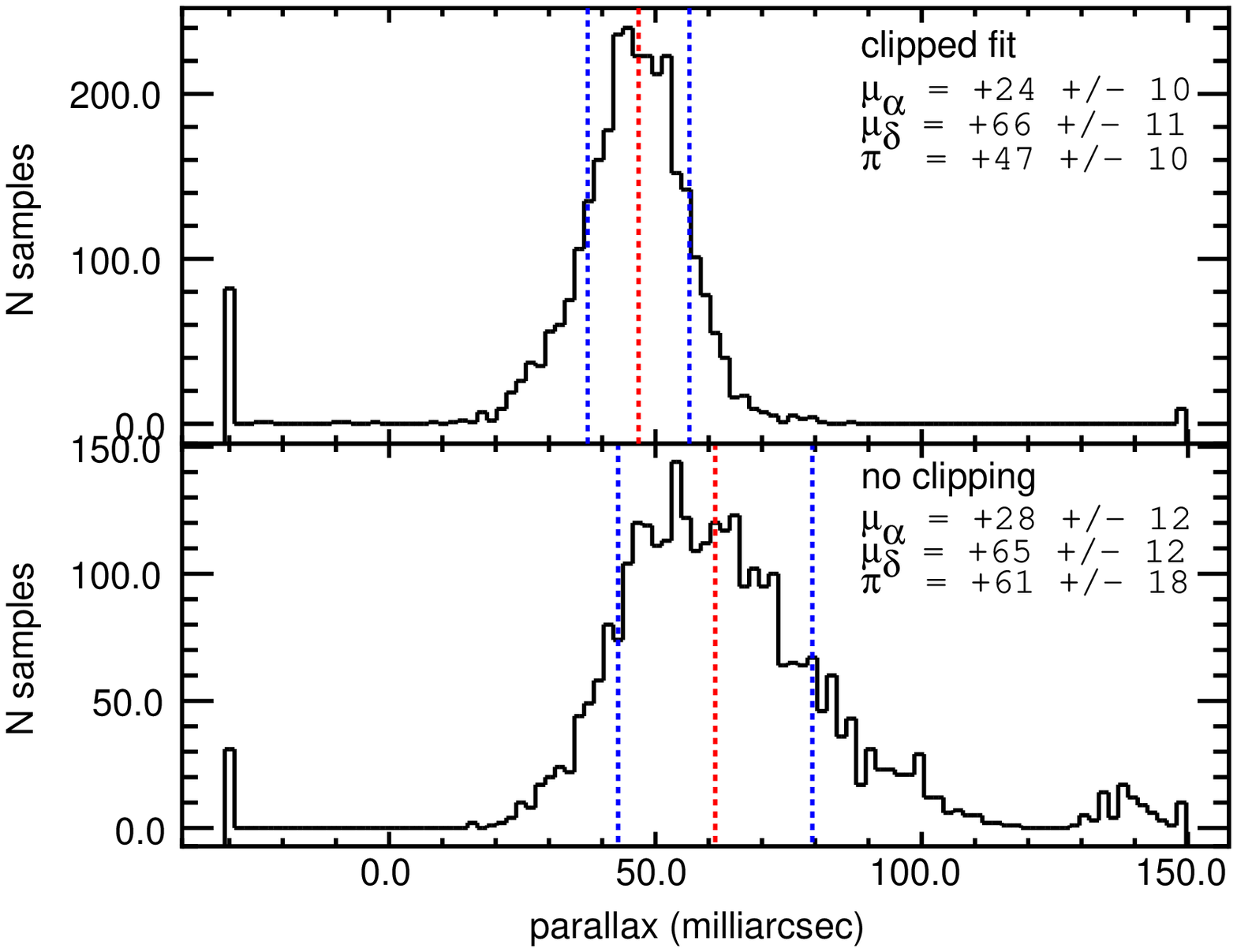}
 \end{picture}
 \caption[]{Bootstrap resampling tests of the astrometric errors.  
{\bf Top:} Histogram of the fitted parallax values for 3000 bootstrap
resampling tests, with all points included.  The blue dashed line
marks the 50\% point in the cumulative distribution function.  The red
dashed lines represent the 15.9\% and 84.1\% points, equivalent to
$\pm 1\sigma$.  The fitted parallax and proper motion values are
listed with the errors derived from this analysis.   
{\bf Bottom:} Histogram of the fitted parallax values for 3000 bootstrap
resampling tests, with the single outlier point excluded.  Labels as above.
} 
  \label{par_hist}
 \end{figure}

Figures~\ref{par_data} and~\ref{par_hist}  show the astrometry of 2MASS
J0213+3648~C.  We have used bootstrap resampling to determine the errors
on the fitted parameters, and to assess whether the inclusion of any specific measurement in the calculation of the fit drives the parameters to significantly different values. For
1000 bootstrap resample tests, we also measured the distance of each
point from the fitted path, scaled by the formal error on the point.
The median of this distribution is a measure of how deviant the point
is from the path given the set of measurements.  We find that one
point, marked with an `X' on Figure~\ref{par_data}, is
significantly deviant from the collection of fitted paths.
The mean parallax changes by a
substantial amount, though formally the two values are within error.
We remove that single discrepant point from the dataset and use the resulting parallax and proper motion values for \objnameshort~C. Our final Pan-STARRS\,1 parallax measurement for \objnameshort~C is 0.045$\pm$0.010\arcsec corresponding to a distance of 22.2$_{-4.0}^{+6.4}$\,pc.

\subsection{WISE photometry}
\begin{figure}
 \setlength{\unitlength}{1mm}
 \begin{picture}(100,60)
 \includegraphics{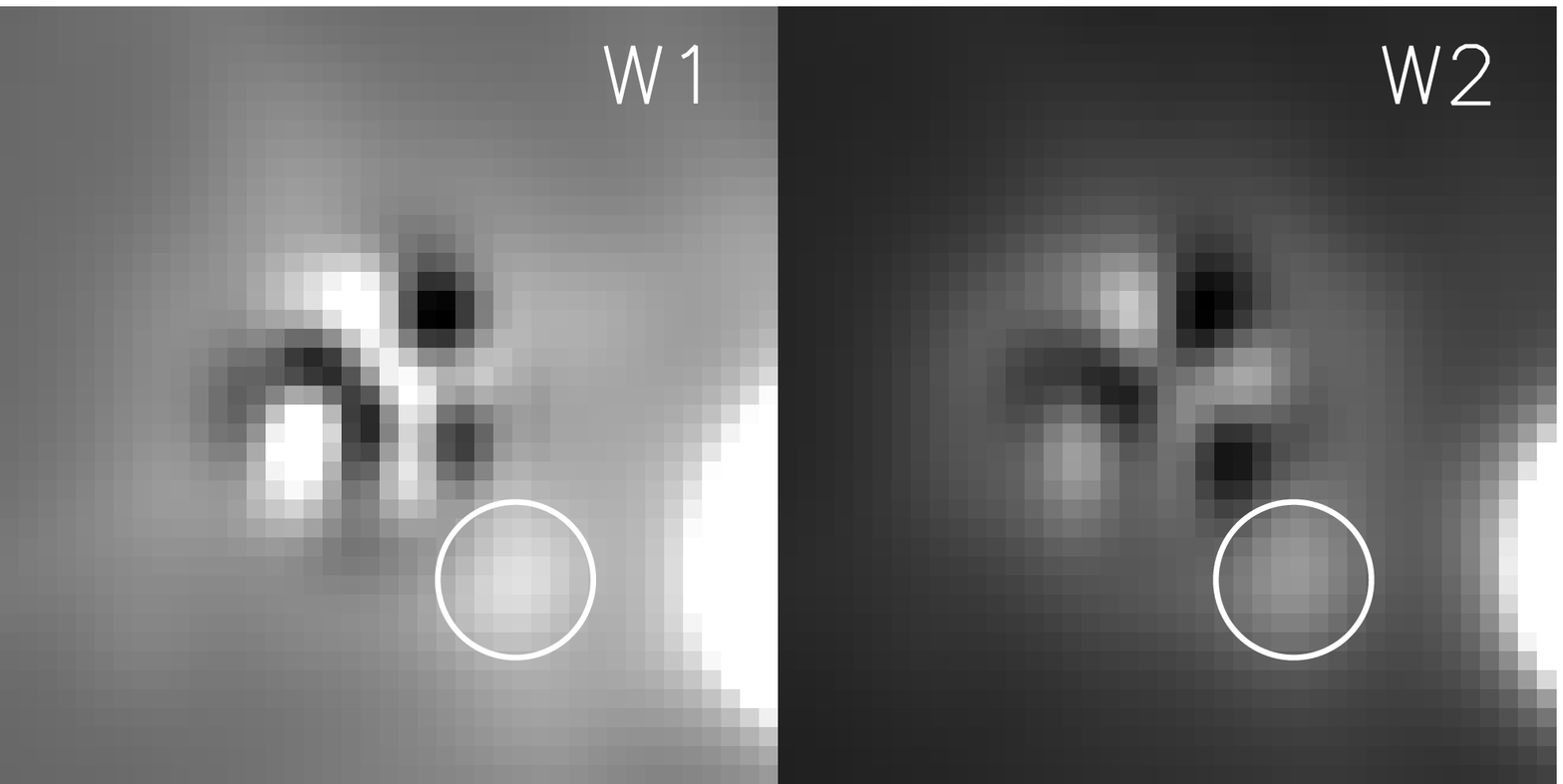}
 \end{picture}
 \caption[]{The results of our WISE PSF subtraction with \protect\objnameshort~C circled. These images are 55 arcseconds across.} 
  \label{psf_sub}
 \end{figure}

\objnameshort~C does not appear in the AllWISE catalogue or reject table, likely due to it's bright primary. To remove the primary and hence estimate the brightness of the C component we modelled the PSF of the primary on downloaded WISE images. \cite{Lang2014} modelled the WISE PSF as three isotropic gaussians. We found that two bivariate gaussians with free axial tilts were a better fit to our primary. The PSF was fitted to the central 10 pixels of the primary to exclude any flux from the secondary. We then subtracted the approximated PSF from the full image revealing the tertiary in better contrast. See Figure~\ref{psf_sub} for the results of this subtraction. Finally we undertook aperture photometry on the tertiary component using two apertures either side of the object but at similar distances from the primary to measure the background flux.  We then used the same technique on other stars in the field and compared the measured fluxes to the WISE catalogue magnitudes in order to determine the image zeropoints. We found that \objnameshort~C had $W_1$=14.204$\pm$0.123\,mag. and $W_2$=13.430$\pm$0.128\,mag.

\section{Results}
The derived properties of the components of the \objnameshort\, system are shown in Table~\ref{prop}. Below we discuss how these parameters were calculated.

\begin{table*}
\begin{minipage}{170mm}
\caption{The properties of the components of the \objname\, system. Note that we do not quote Pan-STARRS\,1 magnitudes for \objnameshort~AB as it is heavily saturated.}

\label{prop}
\begin{center}
\footnotesize
\begin{tabular}{lccc}
\hline
&\multicolumn{2}{c}{\objnameshort\,AB}\\
&\objnameshort\,A&\objnameshort\,B&\objnameshort\,C\\
\hline
Position&\multicolumn{2}{c}{02 13 20.63 +36 48 50.7\,$^{a}$}&02 13 19.82 +36 48 37.5\,$^{a}$\\
$\mu^{\alpha}$ ($''$/yr)&\multicolumn{2}{c}{0.024$\pm$0.008\,$^{e}$}&0.024$\pm$0.010\\
$\mu^{\delta}$ ($''$/yr)&\multicolumn{2}{c}{0.047$\pm$0.008\,$^{e}$}&0.065$\pm$0.011\\
$\pi$ ($''$)&\multicolumn{2}{c}{0.068$\pm$0.020\,$^{e*}$}&0.045$\pm$0.010$^{f}$\\
&&&0.063$^{+0.014}_{-0.010}$\,$^{f*}$\\
Separation&\multicolumn{2}{c}{0.217$''$\,$^{b}$}&16.4$''$\,$^{a}$\\
&\multicolumn{2}{c}{4.8\,AU\,$^{f}$}&360\,AU\,$^{f}$\\
Spectral Type&\multicolumn{2}{c}{M4.5$^c$}&T3\\
&M4.5$^d$&M6.5$^d$&\\
$V$ (mag)&\multicolumn{2}{c}{13.86$^e$}&\ldots\\
$\Delta i$ (mag)&\ldots&2.16$\pm$0.15$^{g}$&\\
$\Delta z$ (mag)&\ldots&2.42$\pm$0.18$^{g}$&\\
$z_{P1}$ (AB mag)&\multicolumn{2}{c}{\ldots}&19.243$\pm$0.016$^{f}$\\
$y_{P1}$ (AB mag)&\multicolumn{2}{c}{\ldots}&17.567$\pm$0.010$^{f}$\\
$J_{2MASS}$ &\multicolumn{2}{c}{9.367$\pm$0.022\,$^{a}$}&15.297$\pm$0.53\,$^{a}$$^\dag$\\
$H_{2MASS}$&\multicolumn{2}{c}{8.825$\pm$0.021\,$^{a}$}&14.765$\pm$0.62\,$^{a}$$^\ddag$\\
$K_{s,2MASS}$&\multicolumn{2}{c}{8.518$\pm$0.018\,$^{a}$}&14.770$\pm$0.115\,$^{a}$$^\ddag$\\
$Y_{MKO}$&\multicolumn{2}{c}{\ldots}&16.279$\pm$0.024$^f$\\
$J_{MKO}$&\multicolumn{2}{c}{\ldots}&15.158$\pm$0.013$^f$\\
$H_{MKO}$&\multicolumn{2}{c}{\ldots}&14.887$\pm$0.021$^f$\\
$K_{MKO}$&\multicolumn{2}{c}{\ldots}&14.930$\pm$0.020$^f$\\
$W_1$&\multicolumn{2}{c}{8.333$\pm$0.023$^{j}$}&14.204$\pm$0.123$^{f,j}$\\
$W_2$&\multicolumn{2}{c}{8.127$\pm$0.020$^{j}$}&13.430$\pm$0.128$^{f,j}$\\
$W_3$&\multicolumn{2}{c}{7.982$\pm$0.021$^{j}$}&\ldots\\
$W_4$&\multicolumn{2}{c}{8.004$\pm$0.229$^{j}$}&\ldots\\
$F_{NUV}/F_J$&\multicolumn{2}{c}{7.1$\times10^{-5}$\,$^{a,f,g}$}&\ldots\\
$F_{FUV}/F_J$&\multicolumn{2}{c}{$<2.2\times10^{-5}$\,$^{f,g,\#}$}&\ldots\\
$F_{X}/F_J$&\multicolumn{2}{c}{5.7$\times10^{-3}$\,$^{a,f,h}$}&\ldots\\
mass (M$_{\odot}$)&0.26$\pm$0.06$^{i}$&0.09$\pm$0.03$^{i}$&0.068$\pm$0.007$^f$\\
age&\multicolumn{2}{c}{1--1.0\,Gyr$\,^{f}$}&\\
\hline
\multicolumn{4}{l}{$^{a}$ 2MASS position, epoch 1998.811 \protect\cite{Skrutskie2006}} \\
\multicolumn{4}{l}{$^{b}$ Epoch 2012.90 \protect\cite{Janson2014}}\\
\multicolumn{4}{l}{$^c$ spectrum of combined object \protect\cite{Riaz2006}}\\
\multicolumn{4}{l}{$^d$ based on flux ratio \protect\cite{Janson2012}}\\
\multicolumn{4}{l}{$^e$ \protect\cite{Lepine2011}}\\
\multicolumn{4}{l}{$^f$ this work} \\
\multicolumn{4}{l}{$^g$ \protect\cite{Janson2012}}\\
\multicolumn{4}{l}{$^h$ \protect\cite{Voges2000}} \\
\multicolumn{4}{l}{$^i$ \protect\cite{Janson2014}}\\
\multicolumn{4}{l}{$^j$ \protect\cite{Wright2010}}\\
\multicolumn{4}{l}{$^\dag$ 2MASS confusion flag set}\\
\multicolumn{4}{l}{$^\ddag$ marked as diffraction spike in 2MASS}\\
\multicolumn{4}{l}{$^*$ photometric parallax}\\
\multicolumn{4}{l}{$^{\#}$ 3$\sigma$ upper limit.}\\
\normalsize
\end{tabular}
\end{center}
\end{minipage}
\end{table*}
\subsection{Properties of \objnameshort~AB}
\objnameshort\, was identified as an M dwarf in the solar neighbourhood with significant X-ray emission by \cite{Riaz2006}. Using low-resolution spectroscopy they classified it as an M4.5 dwarf (a spectral type confirmed by \citealt{Alonso-Floriano2015}) with a spectrophotometric distance of 11\,pc. \cite{Riaz2006}  also reported that the object showed H$\alpha$ emission of 8.1\,\AA.\, \cite{Janson2012} used Lucky Imaging to identify \objnameshort~A as a close binary system with a separation of 0.181$\pm$0.002$"$ and an estimated orbital period of 8 years. We note here that this orbital period is too long for either component of the binary to have its activity significantly affected by tidal spin-up. Subsequent orbital motion combined with the component's flux ratio led \cite{Janson2014} to estimate spectral types of M4.5 and M6.5 and masses of 0.26$\pm$0.06\,M$_{\odot}$ and 0.09$\pm$0.03\,M$_{\odot}$ for the A and B components respectively.

\subsubsection{The age of \objnameshort~AB}
\label{prim_age}
The primary pair have an X-ray counterpart in the {\it ROSAT} Bright Source Catalogue \citep{Voges1999} 13\arcsec\, away, slightly beyond the 1$\sigma$ positional error of 11\arcsec. We estimated the X-ray flux for \objnameshort~AB by applying the relations of \cite{Schmitt1995} to the {\it ROSAT} data yielding a flux of 5.2$\pm1.4\times10^{-13}$\,ergs/cm$^{2}$. We then used 2MASS $J$-band magnitude of \objnameshort~AB to calculate an $F_X/F_J$ flux ratio of 5.7$\times10^{-3}$. Comparing this value to Figure 3 of \cite{Shkolnik2009}, we first note that our target is bluer in $I-J$ than one would expect for an M4.5 (using an $I$ magnitude from SuperCOSMOS; \citealt{Hambly2001}). However this does not affect the conclusion that the object lies on the active M dwarf sequence. This sequence is defined as having similar or greater X-ray emission than the Pleiades (125\,Myr; \citealt{Stauffer1998}) and the $\beta$ Pictoris young moving group (21\,Myr; \citealt{Binks2013}). We note that \objnameshort~AB also emits a higher X-ray flux than most Hyades (625\,Myr; \citealt{Perryman1998}) members of similar spectral type. \cite{Riaz2006} find $\log_{10}L_X/L_{bol}$=3.16, placing it in their saturated X-ray emission locus. Comparing this value with Figure~5 of \cite{Preibisch2005} we find that \objnameshort~AB fractional X-ray flux is similar if not higher than Hyades M~dwarf members and much higher than the vast majority of field M~dwarfs. Using the method of \cite{Malo2014} we calculated that \objnameshort~A/B has $\log L_x=28.5\pm0.6$ergs/s. This is 2$\sigma$ less active than the $\beta$ Pictoris moving group and $1\sigma$ less active than the AB~Dor association. We do however find that it is 1.2$\sigma$ more active than the typical field star of similar type.

As our object is a binary it is possible that the X-ray flux is coming entirely from the M6.5 secondary component. This would not alter our conclusion that the object is X-ray active as a recalculation of $F_X/F_J$ would result in the object lying even further above the inactive M dwarf regime on Figure 3 of \cite{Shkolnik2009}. If we were relying solely on X-ray emission for an age diagnostic, we would estimate that this object is younger than the Hyades (i.e. $<$625\,Myr).

\objnameshort~AB are also detected by the {\it GALEX} satellite \citep{Martin2005} in the near-UV band but not in the far-UV band. We note that it does appear as a far-UV emitter in the EUVE catalogue \citep{Christian1999} due to a stellar flare. The {\it GALEX} near-UV flux of 20.1$\pm$2.4~$\mu$Jy results in an $F_{NUV}/F_J$ ratio of 7.1$\times10^{-5}$ or an $F_{NUV}/F_{K_s}$ ratio of 7.7$\times10^{-5}$. We note that the latter value is within 0.2\,dex of the young (20--125\,Myr) M4.5 binary TYC~5241-986-1~BC \citep{Deacon2013}. We set an upper limit on the FUV flux of \objnameshort~AB by finding the flux error on the forced far-UV photometry at the position of the {\it GALEX} NUV detection. From this we calculate a 3$\sigma$ upper limit of $F_{FUV}/F_J <2.2\times10^{-5}$. To further examine the near-UV properties of \objnameshort~AB, we must compare with Figure~3 of \cite{Shkolnik2011}. Here we find that our object appears to lie between the active and inactive loci and that while it does not have a far-UV, detection it could have emission above \cite{Shkolnik2009}'s $F_{NUV}/F_J>10^{-5}$ and still remain undetected by {\it GALEX}. However the near-UV emission is of little use as \cite{Ansdell2014} noted that the fully convective transition in mid-M~dwarfs makes it impossible to define a near-UV inactive locus of M~dwarfs beyond M3. Hence we do not consider \objnameshort~AB's near-UV emission to be a reliable age diagnostic.

\cite{Riaz2006} found that \objnameshort~AB has H$\alpha$ in emission. \cite{West2008} provide a list of activity lifetimes as a function of spectral type. As the activity lifetime changes with spectral type we need to know which component this emission comes from. While we were not able to resolve the H$\alpha$ emission into two separate components in our CAFE spectrum, we were able to measure its equivalent width as $-$6.6\,\AA,\, broadly in agreement with \cite{Riaz2006}'s value of $-$8.1\,\AA\, and the value of $-6.2^{+0.4}_{-0.2}$\,\AA\, from \cite{Alonso-Floriano2015}. This emission could in theory come from either component of the inner binary. Using the mean $\chi$ correction factors as a function of spectral type of \cite{Walkowicz2004} we estimated the $\log L_{H\alpha}/L_{bol}$ to be -3.4 (from our EW of -6.6\AA)\, if the emission was coming from the A component. This is in the most active quartile for stars of a similar spectral type in Figure~6 of \cite{Schmidt2015} but is not outside the bounds of the activity range in their sample. For the secondary component we used the $i$ band flux ratio of \cite{Janson2012} and the typical $r-i$ colour difference between an M4.5 and M6.5 in the SED templates of \cite{Kraus2007} to estimate an $r$-band contrast ratio of 2.59. We then used this value to adjust the EW of the H$\alpha$ feature to take into account the lower continuum in the $r$-band for an M6.5. This resulted in an approximate equivalent width of 72\AA\, if the emission is coming entirely from the secondary. Again using the method of  \cite{Walkowicz2004} we calculated $\log L_{H\alpha}/L_{bol}$ to be -3.0 if all the H$\alpha$ emission was coming from the secondary component. This is more active than any M6 or M7 in \cite{Schmidt2015}'s Figure~6 (with $\log L_{H\alpha}/L_{bol}$ declining with increasing spectral type). Hence we conclude that it is unlikely that all the H$\alpha$ emission is from the secondary component and that the M4.5 A component must be active. However \cite{Morgan2012} show that binaries in the 1--100\,AU separation range have significantly higher activity fractions. Hence we cannot use this activity to set an upper age bound. 

Young M~dwarf stars in star forming regions typically have infrared excesses caused by circumstellar disks. We used the WISE photometry \citep{Wright2010} for \objnameshort~AB to examine if it has mid-infrared excess indicative of a disk. Comparing with the $W_3-W_4$ vs. $W_1-W_2$ plot shown in Figure~1 of \cite{Simon2012} we find that our target falls close to the stellar locus suggesting that our primary does not host a disk. As discussed in \cite{Deacon2013}, disks around M~dwarfs are uncommon for ages beyond 20\,Myr. While studies of younger (10--15\,Myr) clusters show significant amounts of disk emission around many M~dwarfs (e.g. \citealt{Currie2008}), \cite{Simon2012} found no disk excesses in the M~dwarf population of Tucana Horologium  (45$\pm$4\,Myr, \citealt{Bell2015}) and AB~Dor (149$^{+51}_{-19}$ \citealt{Bell2015}). Hence the lack of disk excess indicates that \objnameshort~AB is older than around 20\,Myr.

\begin{figure}
 \setlength{\unitlength}{1mm}
 \begin{picture}(100,70)
 \includegraphics{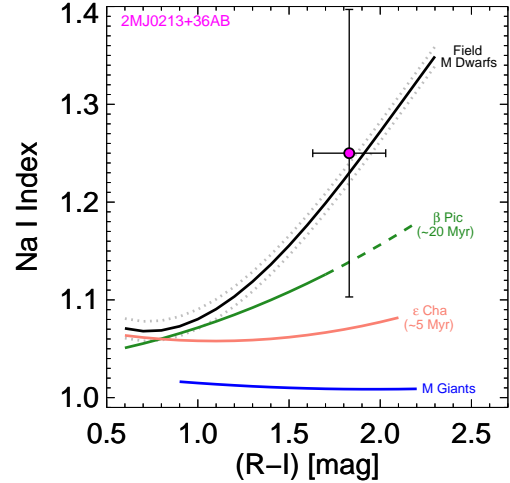}
 \end{picture}
 \caption[]{The NaI index from \protect\cite{Lyo2004} along with the measured sequences for different populations from \protect\cite{Lawson2009}. \protect\objnameshort\, lies just above the old field sequence.} 
  \label{Lyo_ind}
 \end{figure}

The Na 8200\,\AA~ doublet provides an effective gravity diagnostic for mid-late M dwarfs \citep{Lyo2004,Schlieder2012a}. We measured the equivalent width of this feature in our IRTF spectrum and found it to be 5.8$\pm$1.8\,\AA,~on the upper end of the field dwarf sequence in \cite{Schlieder2012a}'s Figure~3. This points towards an older lower age boundary of $\sim$1\,Gyr. A visual comparison with the young (20--125\,Myr) M4.5 TYC~5241-986-1~B/C \citep{Deacon2013} also indicates that \objnameshort~AB is not a young object (see Figure~\ref{prispec} inset). We also used the NaI spectral index derived by \cite{Lyo2004} and measured a value of 1.25$\pm$0.15. Figure~\ref{Lyo_ind} shows that our object has a feature strength (and thus a gravity) consistent with field dwarfs. While the M6.5 component will have a deeper 8200\,\AA~doublet owing to its late spectral type, the contrast ratio of $\Delta i$ =2.16$\pm$0.15\,mag indicates that it will contribute less than 10\% of the flux in this region and thus will not significantly change the line depth. This NaI spectral index value has a relatively high error due to low S/N in the surrounding continuum region. Both the visual comparison of the 8200\AA~feature and the lack of other youth indicators suggest that this is not a young object. Additionally we found that the strength of TiO (TiO5 0.436, M3.7; TiO6 0.66, M4.0), CaH (CaH3 0.61, M5.6), VO (VO1 0.90, M5.4; VO2 0.72, M4.9) and H$_2$O (H2OD 1.14) molecular indices in the optical \citep{Lepine2013} and near-infrared \citep{McLean2003,Allers2013} absorption features of \objnameshort~A/B are consistent with a spectral type of M5.0$\pm$0.5. This is in agreement with the literature spectral type of M4.5 \citep{Riaz2006}.

\begin{figure*}
 \setlength{\unitlength}{1mm}
 \begin{picture}(100,120)
 \includegraphics{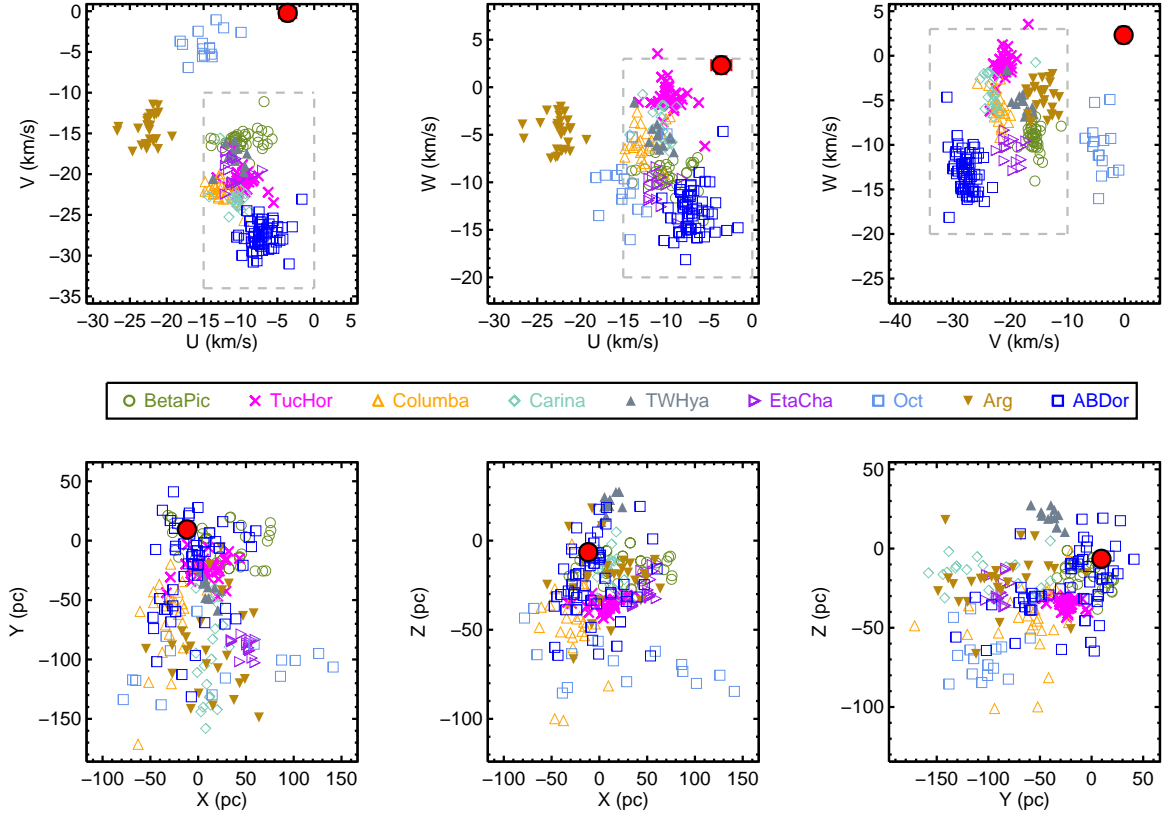}
 \end{picture}
 \caption[]{A plot showing the positions of known young moving groups and our target system \objnameshort\, (red circle). The dotted box shows the broadly young Local Association \protect\citep{Zuckerman2004}.} 
  \label{ymg}
 \end{figure*}

One final youth indicator available to us is kinematics. We ran \objnameshort~AB through the BANYAN~II online tool \citep{Malo2013,Gagne2014} and found that its kinematics do not match any of the known young moving groups. We also plotted our object on a series of diagnostic plots comparing its Galactic XYZ positions and UVW space velocities (Figure~\ref{ymg}). Again we find that \objnameshort~AB does not match any known young moving groups and lies outside the loosely defined young ($\la$125\,Myr) Local Association \citep{Zuckerman2004}. We also found that \objnameshort~AB did not match the kinematics of other groups not shown in our plot; Octans \citep{Murphy2014b}, Hercules-Lyra \citep{Eisenbeiss2013}, 32 Ori \citep{Mamajek2006}, Carina Near \citep{Zuckerman2006} and Ursa Majoris \citep{King2003}.

In summary we assign an upper age bound of 10\,Gyr for \objnameshort~AB based on disc kinematics and a lower age boundary of 1\,Gyr based on the strength of the Na 8200\,\AA~feature. We note that \objnameshort~AB has X-ray emission indicative of it being younger than the Hyades. However we caution that the relatively late spectral type of this object and its multiplicity make it difficult to apply some age diagnostics such as UV emission with sufficient certainty.

\subsection{Properties of \objnameshort~C}
\label{spec_an}
\begin{figure}
 \setlength{\unitlength}{1mm}
 \begin{picture}(100,70)
 \includegraphics{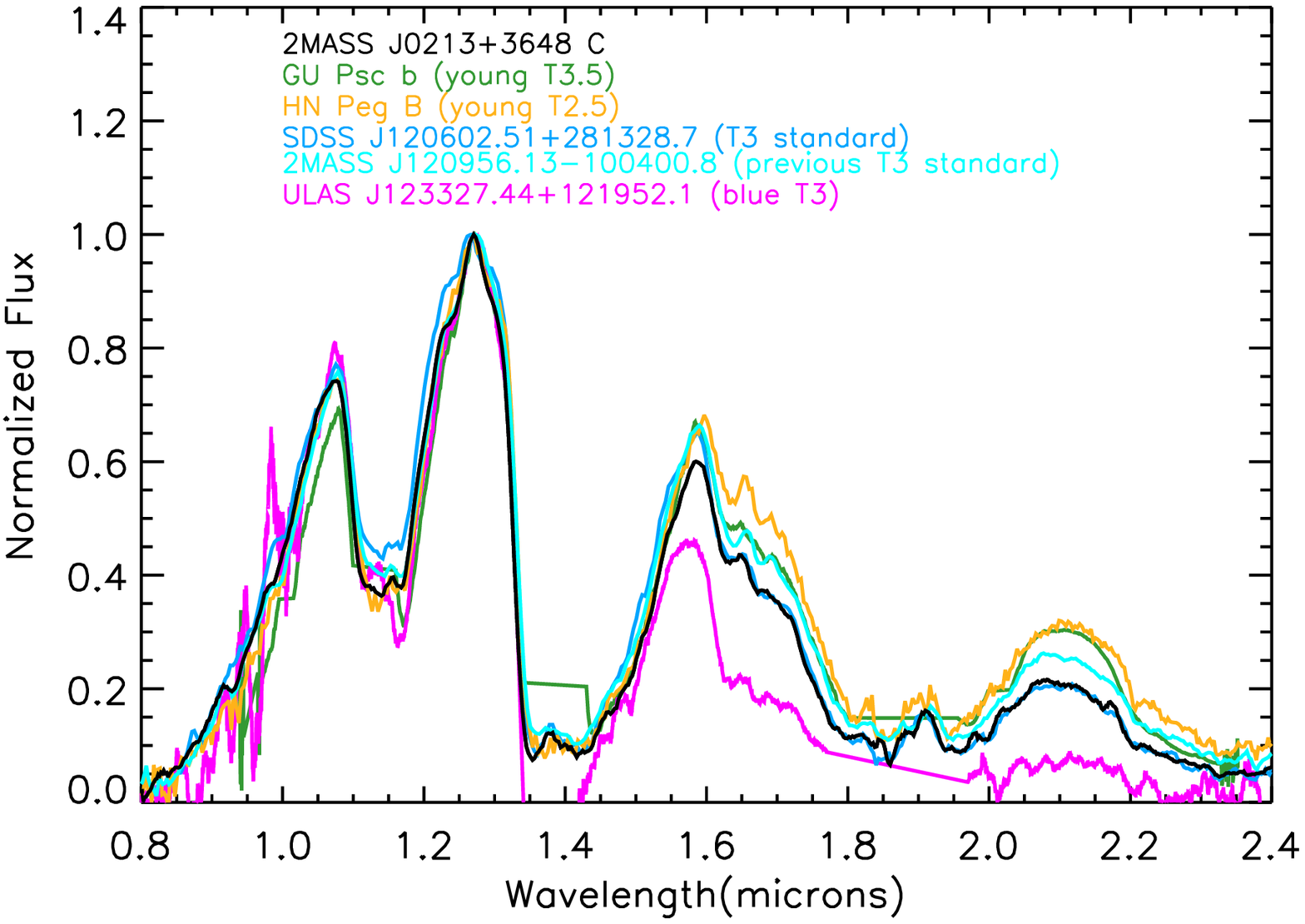}
 \end{picture}
 \caption[]{Our spectrum of \objname~C, we classify this is object as a T3$\pm$0.5. Shown for comparison are the spectra for the young ($\sim$100\,Myr) T3.5 GU~Psc~b \protect\citep{Naud2014}, the young ($\sim$300\,Myr) T2.5 HN~Peg~B \citep{Luhman2007}, the alternative T3 dwarf standard SDSSJ120602.51+281328.7 \protect\citep{Chiu2006} (as suggested by \protect\citealt{Liu2010}), the previous T3 standard 2MASS~J120956.13$-$100400.8 (\citealt {Burgasser2004,Burgasser2006}; now known to be a binary) and the blue T~dwarf \citep{Burningham2010,Marocco2015}. All spectra are smoothed to similar resolutions and normalised to the $J$-band peak.} 
  \label{sec_spectrum}
 \end{figure}

\begin{figure*}
 \setlength{\unitlength}{1mm}
 \begin{picture}(100,130)
 \includegraphics{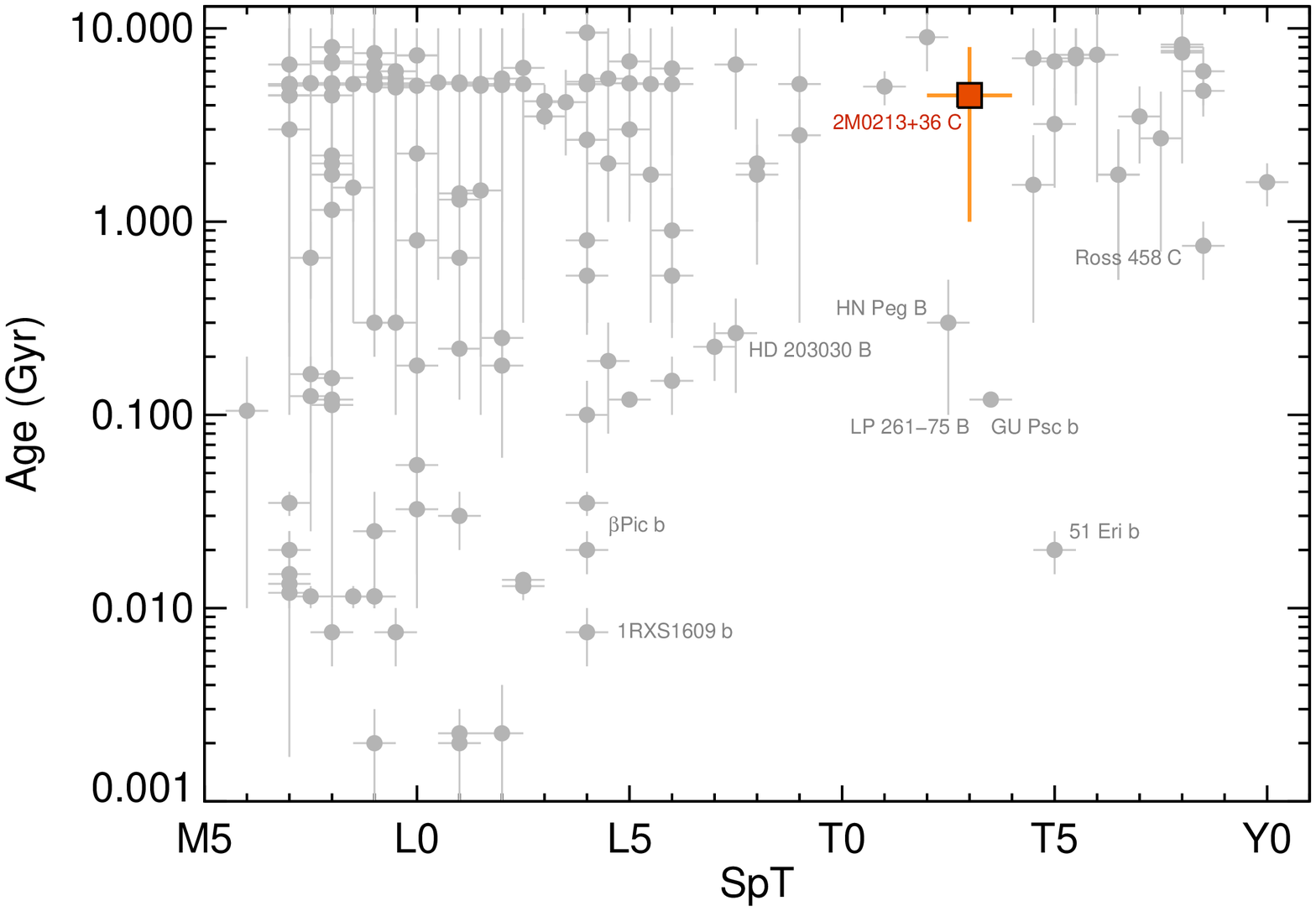}
 \end{picture}
 \caption[]{An updated version of the benchmark companion age-spectral type plot from \protect\cite{Bowler2015}. The data for this plot comes largely from the compilation of \protect\cite{Deacon2014}. \protect\objnameshort~C adds a field-age early-mid T benchmark into and area of the diagram which was previously sparsely populated.} 
  \label{age_spt}
 \end{figure*}

The observed near-infrared spectrum of \objnameshort~C is shown in Figure~\ref{sec_spectrum}. We spectrally classified \objnameshort~C using the flux indices of \cite{Burgasser2006} and the polynomial relations of \cite{Burgasser2007}. The individual indices and the derived spectral types are shown in Table~\ref{spec_ind}. We also compared visually with the standards of \cite{Burgasser2006}, finding a best fit of T3 (note the T3 standard of \citealt{Burgasser2006} was found to be a binary by \citealt{Liu2010}, see later discussion on the comparison with the newer standard). As this compares well with the spectral types derived from the indices we adopt this as our final spectral type. Figure~\ref{sec_spectrum} also shows a comparison with the young ($\sim100$\,Myr) T3 GU~Psc~b \citep{Naud2014}, the young ($\sim$300\,Myr) T2.5 HN~Peg~B \citep{Luhman2007}, the original T3 standard (2MASS~J120956.13$-$100400.8, \citealt{Burgasser2006}) and the alternative T3 standard (SDSS~J120602.51+281328.7) suggested by \cite{Liu2010} after 2MASS~J120956.13$-$100400.8 was found to be a binary. We note that \objnameshort~C has a narrower $J$ band peak and deeper $J$-band water absorption than SDSS~J120602.51+281328.7 but matches both HN~Peg~B and 2MASS~J120956.13$-$100400.8 well in this regime, the $H$ and $K$-bands the spectrum are a better fit to SDSS~J120602.51+281328.7. GU~Psc~b is unique amongst our comparison objects in that it shows a distinct slope in the $Y$ and $J$-bands compared to the other objects. We do not see sufficient evidence to suggest that \objnameshort~C is spectrally peculiar and note that it does not have the enhanced $K$-band emission that \cite{Naud2014} attributed to reduced collision induced absorption resulting from the lower surface gravity (i.e. younger age) of GU~Psc~b.  \objnameshort~C is the only known T3 benchmark companion known (see Figure~\ref{age_spt}). Our measured WISE photometry yields $W_1-W_2$=0.77, in line with the expect colour for an early-mid T dwarf \citep{Kirkpatrick2011}.

\begin{table*}
\begin{minipage}{170mm}
\caption{Spectral classification of \objnameshort~C. Shown are the spectral indices \citep{Burgasser2006} and the implied spectral classifications from polynomial relations \citep{Burgasser2007} for each index. }

\label{spec_ind}
\begin{center}
\tiny
\begin{tabular}{ccccccc}
 \hline
H$_2$O-J (SpT)&CH$_4$-J (SpT)&H$_2$O-H (SpT)&CH$_4$-H (SpT)&CH$_4$-K (SpT)&Visual&Final\\
&&&&&Type&Type\\
\hline
0.395$\pm$0.013 (T3.7)&0.450$\pm$0.019 (T4.3)&0.435$\pm$0.026 (T3.7)&0.709$\pm$0.027 (T2.5)&0.432$\pm$0.034 (T3.1)&T3&T3\\
\hline
\normalsize

\end{tabular}
\end{center}
\end{minipage}
\end{table*}

To compare with our trigonometric parallax measurement from Pan-STARRS\,1, we calculated a photometric distance for \objnameshort~C using only our WFCAM photometry. This was done in a similar manner to \cite{Deacon2014} using the relations of \cite{Dupuy2012}, by calculating distances in each band along with the error on these distances caused by photometric measurement error and the error in spectral type. We then used the quadrature sum of these error terms to estimate a distance based on the J, H \& K bands. Our final distance is 15.08$^{+3.0}_{-2.5}$\,pc which includes both the calculated error due to photometric and spectroscopic measurement uncertainty and the scatter around the absolute magnitude relations. It compares well with the photometric distance of the AB component 14.7$^{+6.2}_{-3.3}$\,pc \citep{Lepine2011} and the parallactic distance of 22.2$^{+6.3}_{-4.0}$\,pc.

\subsubsection{Comparison with evolutionary models}
\begin{figure}
 \setlength{\unitlength}{1mm}
 \begin{picture}(100,90)
 \includegraphics{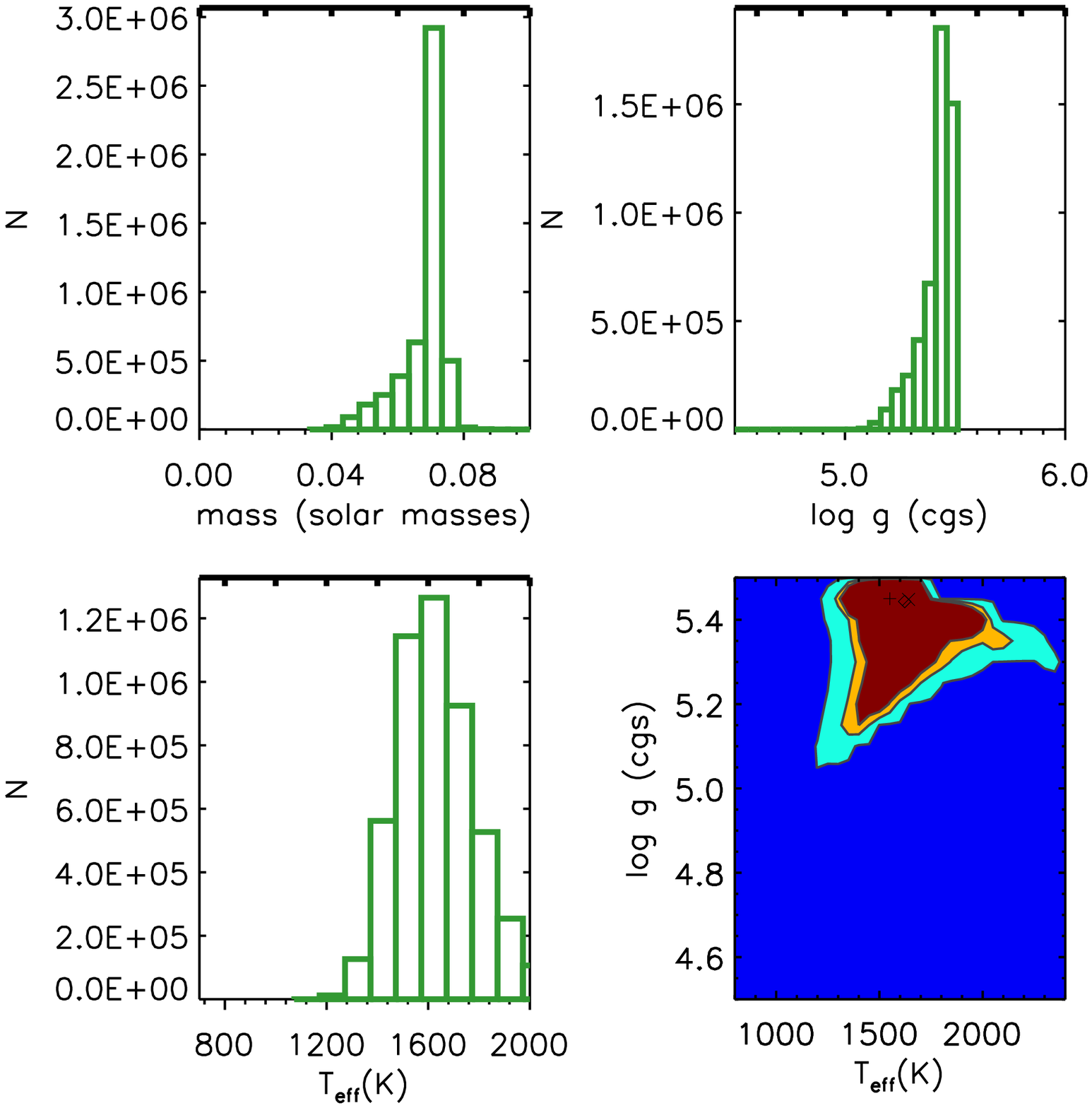}
 \end{picture}
 \caption[]{Evolutionary model predictions for \protect\objnameshort~C assuming a flat age range of 1--10\,Gyr. The results of 5,000,000 Monte Carlo realisations of the object and comparison with the \protect\cite{Baraffe2003}. The $+$ symbol in the bottom right panel marks the mean $T_{eff}$ and $\log g$ predicted by the comparison and the diamond marks the median values while the cyan, yellow and maroon regions show the 3, 2 and 1 $\sigma$ confidence regions respectively.} 
  \label{ev_mod_fig}
 \end{figure}

We calculated the bolometric correction for \objnameshort~C using the polynomial relations of \cite{Liu2010}. This resulted in an MKO J-band correction of $BC_J$=2.24$\pm$0.14. We combined this with our WFCAM observations to yield an apparent bolometric magnitude of 17.40$\pm$0.50\,mag (including the error in our parallax measurements). To determine the physical properties of \objnameshort~C, we ran a Monte Carlo comparison to the \cite{Baraffe2003} evolutionary models. An age for the system was drawn from a flat distribution with between our upper and lower age boundaries (10\,Gyr and 1\,Gyr) and the apparent bolometric magnitude and parallax were both offset by a random gaussian offset with a standard deviation equal to the measurement errors. This resulted in an age and absolute bolometric magnitude for each realisation. The mass, effective temperature and gravity corresponding to that absolute bolometric magnitude and age were then determined from the \cite{Baraffe2003} model grid. The results are shown in Figure~\ref{ev_mod_fig} with values of $T_{eff}=1641\pm167$\,K, $m=68\pm7$M$_{J}$ and $\log g = 5.45 \pm 0.08$\,dex (cgs). Note this relatively high temperature (for a T3) is likely due to our distance from the trigonometric parallax being larger than our photometric distance.

\subsubsection{The absolute magnitude of \objnameshort~C}
We calculated absolute magnitudes in the MKO system based on both our trigonometric parallax for the tertiary component and the \cite{Lepine2011} photometric parallax for the unresolved inner pair. These are for the trigonometric parallax $M_J= 13.424\pm0.483$\,mag, $M_H= 13.153\pm0.483$\,mag \& $M_K= 13.196\pm0.483$\,mag and $M_J= 14.321\pm0.639$\,mag, $M_H= 14.050\pm0.639$\,mag \& $M_K= 14.093\pm0.639$\,mag with the error estimates dominated by the uncertainties on the distance measurements. Comparing with the absolute magnitude to spectral type plots of Figure~27 of \cite{Dupuy2012} we find that the absolute magnitudes calculated with the photometric parallax lie on the main locus of early T dwarfs. The absolute magnitudes based on the trigonometric parallax are approximately one magnitude over-luminous or approximately 2$\sigma$. This suggests that either the trigonometric parallax is a chance underestimate or perhaps that the secondary is over-luminous due to binarity. Both of these explanations would also fit our anomalously hot effective temperature estimate from evolutionary models.

 \subsubsection{Comparison with atmospheric models}
\begin{figure*}
 \setlength{\unitlength}{1mm}
 \begin{picture}(100,80)
 \includegraphics{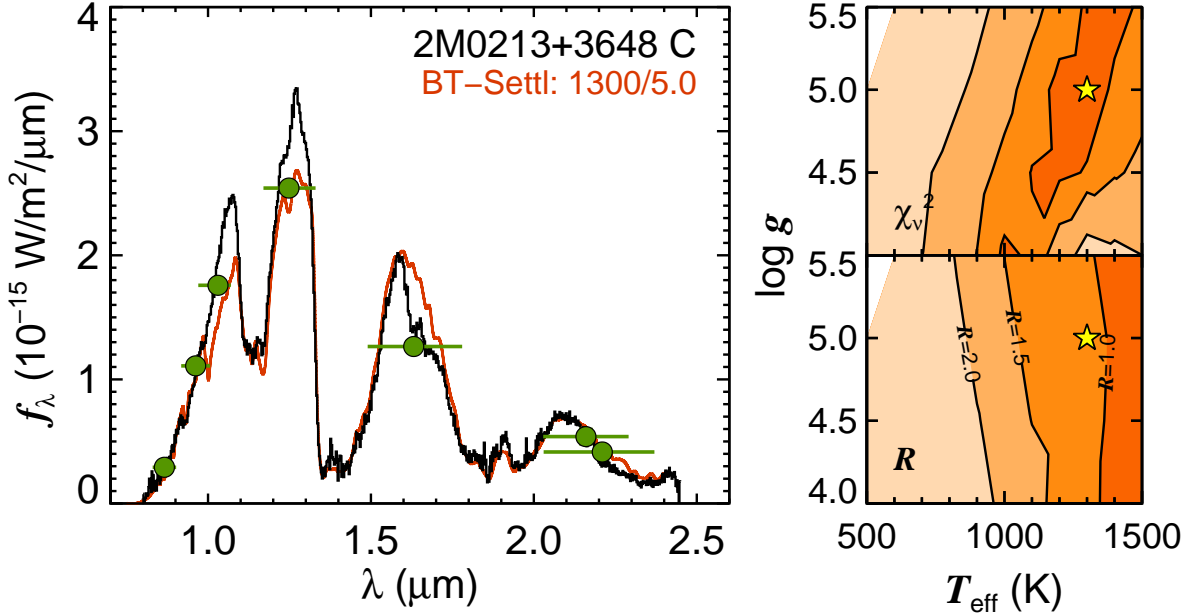}
 \end{picture}
 \caption[]{The result of our comparison between our spectrum for \protect\objnameshort~C and the BT-Settl models \protect\citep{Allard2010}. The green points in the left-hand panel show our $Y$, $J$, $H$ and $K$ UKIRT photometry and the 2MASS $K_S$ photometry. The right-hand panel show the likelihood contours for the different parameters with the most probable values marked with a star.} 
  \label{specfit}
 \end{figure*}

Atmospheric models provide an independent way to infer the physical properties of \objnameshort~C.  We fit the solar metallicity BT-Settl-2010 grid of synthetic spectra \citep{Allard2010} spanning
500--1500\,K in effective temperature and 4.0-5.5~dex in surface gravity (log~$g$ [cgs]) to our prism spectrum, flux-calibrated to our UKIRT $J$-band photometry. The 1.60--1.65~$\mu$m region is ignored in the
fits owing to poorly calibrated methane opacities in this region of the models.  The 1.80--1.95~$\mu$m region is also excluded because telluric absorption is particularly strong and difficult to accurately
correct in this region.  The fitting procedure follows a maximum likelihood approach as described in detail in \cite{Bowler2011}. In brief, we calculate reduced $\chi^2$ values for each model in 100~K increments in $T_\mathrm{eff}$ and 0.5~dex in log~$g$.  The multiplicative factor to scale the model to the flux-calibrated spectrum corresponds to the ratio of the object's radius to its distance, squared.  Adopting the astrometric distance of 22~pc yields a corresponding radius for each synthetic spectrum.  The best fit model has $T_\mathrm{eff}$=1300~K, log~$g$=5.0~dex, and a radius of 1.1~$R_\mathrm{Jup}$ (Figure~\ref{specfit}).  This model does not reproduce our spectrum very well, however, particularly in the 1.0--1.3~$\mu$m range. Systematic errors of several hundred Kelvin are not uncommon when fitting model atmospheres to brown dwarf spectra as a result of incomplete line lists, missing sources of opacity, and non-solar chemical abundances (e.g. \citealt{Deacon2012}).

\section{Conclusions}
We have discovered a wide T3 companion to the M4.5+M6.5 binary \objname. T dwarfs in higher-order multiple systems are relatively rare, this being the fifth such system discovered after Gl 570 D \citep{Burgasser2000},Wolf 1130 B \citep{Mace2013}, Ross 458 C \citep{Goldman2010} and $\xi$ UMa E, \citep{Wright2013}. The central binary of \objnameshort\, has H$\alpha$ and X-ray activity along with UV flaring suggesting it may be young (with the X-ray emission pointing towards it being younger than the Hyades). However its strong Na 8200\,\AA\, absorbtion and kinematics inconsistent with known young moving groups means we adopt a conservative age range of 1--10\,Gyr. This means that \objname~C provides an older equivalent to the young, similar spectral type benchmarks GU~Psc~b \citep{Naud2014} and HN~Peg~B \citep{Luhman2007}. Figure~\ref{age_spt} shows how our object fills a previous gap in field-age early-mid T benchmark systems.

\section*{Acknowledgments}
The Pan-STARRS1 Surveys (PS1) have been made possible through contributions of the Institute for Astronomy, the University of Hawaii, the Pan-STARRS Project Office, the Max-Planck Society and its participating institutes, the Max Planck Institute for Astronomy, Heidelberg and the Max Planck Institute for Extraterrestrial Physics, Garching, The Johns Hopkins University, Durham University, the University of Edinburgh, Queen's University Belfast, the Harvard-Smithsonian Center for Astrophysics, the Las Cumbres Observatory Global Telescope Network Incorporated, the National Central University of Taiwan, the Space Telescope Science Institute, the National Aeronautics and Space Administration under Grant No. NNX08AR22G issued through the Planetary Science Division of the NASA Science Mission Directorate, the National Science Foundation under Grant No. AST-1238877, the University of Maryland, and Eotvos Lorand University (ELTE) and the Los Alamos National Laboratory. This research has benefitted from the SpeX Prism Spectral Libraries, maintained by Adam Burgasser at http://pono.ucsd.edu/~adam/browndwarfs/spexprism . This publication makes use of data products from the Two Micron All Sky Survey, which is a joint project of the University of Massachusetts and the Infrared Processing and Analysis Center/California Institute of Technology, funded by the National Aeronautics and Space Administration and the National Science Foundation. This publication makes use of data products from the Wide-field Infrared Survey Explorer, which is a joint project of the University of California, Los Angeles, and the Jet Propulsion Laboratory/California Institute of Technology, funded by the National Aeronautics and Space Administration. This publication also makes use of data products from NEOWISE, which is a project of the Jet Propulsion Laboratory/California Institute of Technology, funded by the Planetary Science Division of the National Aeronautics and Space Administration. Based on observations made with the NASA Galaxy Evolution Explorer. {\it GALEX} is operated for NASA by the California Institute of Technology under NASA contract NAS5-98034. The authors would like to thank the Mike Irwin and Simon Hodgkin at the Cambridge Astronomical Survey Unit for making reduced WFCAM data available. The {\it ROSAT} project was supported by the Ministerium f{\"u}r Bildung, Wissenschaft, Forschung und Technologie (BMBF/DARA) and by the Max-Planck-Gesellschaft. The authors wish to recognize and acknowledge the very significant cultural role and reverence that the summit of Mauna Kea has always had within the indigenous Hawaiian community. We are most fortunate to have the opportunity to conduct observations from this mountain. This research made use of TOPCAT, an interactive graphical viewer and editor for tabular data \citep{Taylor2005}. Partial support for this work was provided by National Science Foundation grants AST-1313455 and AST-0709460.

\bibliography{ndeacon}

\bibliographystyle{mn2e}
\label{lastpage}

\end{document}